\documentclass{eas}
\usepackage{graphicx}
\usepackage{natbib}
\usepackage{hyperref}
\usepackage[]{hyperref}
\hypersetup{colorlinks=true,citecolor=blue}
\usepackage{fixltx2e}
\begin{document}

\title{Main ways in which stars influence the climate and surface habitability of their planets} 
\author{Martin Turbet}\address{Observatoire Astronomique de l'Universit{\'e} de Gen{\`e}ve, Chemin Pegasi 51, 1290 Sauverny, Switzerland}
\author{Franck Selsis}\address{Laboratoire d’astrophysique de Bordeaux, Univ. Bordeaux, CNRS, B18N, Allée Geoffroy Saint-Hilaire, F-33615 Pessac, France}
\begin{abstract}
We present a brief overview of the main effects by which a star will have an impact (positive or negative) on the surface habitability of planets in orbit around it. Specifically, we review how spectral, spatial and temporal variations in the incident flux on a planet can alter the atmosphere and climate of a planet and thus its habitability. For illustrative purposes, we emphasize the differences between planets orbiting solar-type stars and late M-stars. The latter are of particular interest as they constitute the first sample of potentially habitable exoplanets accessible for surface and atmospheric characterization in the coming years.
\end{abstract}
\maketitle
\section{Introduction}

From birth to death, the evolution of the nature of a planet (i.e. of its interior, surface and atmosphere) is intrinsically linked to that of its host star. Over time, as the nature of a planet evolves, its ability to offer all the necessary conditions for the emergence and development of life will thus evolve.

Based on our experience on Earth, we can identify some necessary (and certainly not sufficient) conditions  for life to emerge and develop such as the availability of elements to make complex organic molecules (the famous "CHNOPS": carbon, hydrogen, nitrogen, oxygen, phosphorus and sulfur).
Liquid water is considered an irreplaceable solvent for habitability (see \cite{Forget:2013drake} and references therein) but while it is known to exist in the interior of a large variety of planetary bodies (Europa, Ganymede, Callisto, Titan, Pluto, Triton, etc. ; see \cite{Lunine:2017} and references therein), planets that have liquid water on their surface have additional qualities. In fact, a source of low-entropy energy to initiate the irreversible chemical reactions necessary to initiate proto-metabolisms may require the exposition of liquid water to stellar radiations (\cite{Boiteau&Pascal:2011,2016:Pascal}). In the case of exoplanets, which will remain for a long time out of reach for in-situ exploration, identifying signs of life from interstellar distances (if possible) is likely to require photosynthetic processes and therefore surface liquid water. Lithoautotrophic life able to thrive without stellar light may indeed be unable to modify a whole planetary environment in an observable way (\cite{Rosing2005,Kasting:2014}). In summary, the remote search for habitable planets outside the solar system boils down to the search for planets capable of maintaining liquid water on their surfaces (\cite{Kasting:1993,Kopparapu:2013}). This is why we focus in this manuscript on the main effects by which a star will have an impact on the ability of its planets to maintain surface liquid water.

\medskip

A first naive thing that comes to mind when thinking of the effects that a star may have on the habitability of its planets is the following thought experiment: what would happen to a planet without a star? Science fiction provides ideas of the situations in which this could happen. Although a planet around a black hole seems like an attractive configuration (see e.g., the \textit{Interstellar} movie), black holes may lead to several warming mechanisms (e.g., by the blueshifting and beaming of incident radiation of the background stars on the planet ; see \cite{Schnittman:2019}) so that the system is not equivalent to a planet with no star.
Better analogues are more likely rogue or free-floating planets, which are wandering planets without star. We now have direct evidence that such planets do exist by collecting the light directly emitted by the youngest, hottest and thus brightest of these objects (\cite{Zapatero:2000,Luhman:2005,
Bihain:2009,Delorme:2012,Liu:2013,Dupuy:2013}).
Although these direct detections are limited so far to massive planets (several times the mass of Jupiter), we know these rogue planets exist in the terrestrial-mass regime too thanks to micro-lensing surveys (\cite{Sumi:2011,Mroz:2017}). The most likely formation scenario for these low-mass planets is that they have been ejected from their initial planetary system (see \cite{Laskar:1994,Kroupa:2003,Ma:2016} and references therein).
Rogue planets can maintain a subsurface ocean by the insulation of their internal heat flux by a thick layer of, for example, ice (\cite{Abbot:2011}). They can even keep a surface temperature above 273~K under a dense enough H$_2$ atmosphere (\cite{Stevenson:1999,Seager:2013}) thanks to the opacity induced by H$_2$-H$_2$ collisions  (\cite{Borysow:2002,Pierrehumbert:2011h2,Ramirez:2017b}). However, in the absence of stellar irradiation and given the challenge of sounding their atmosphere, the habitability of these objects is likely to remain an academic case for some time, unless there is a close encounter with the solar system (\cite{Abbot:2011}).
In short, having a star seems to be a necessary ingredient for a planet to be potentially habitable in a detectable way.

\medskip

Therefore, we present below a brief overview of the dominant effects by which a star will have an impact on the habitability of planets in orbit around it. 
A more general list of all the effects that can affect the habitability of planets is provided in \cite{Segura:2018} and \cite{Meadows:2018} but the main influence comes from the type of the host star, which strongly impacts both the possible lifespan of habitable conditions and our ability to remotely probe planetary environments. Stars more massive than 2.5~M$_\odot$ live less than 1~Gyr and during this short life their luminosity increases dramatically offering conditions compatible for life on their planets for a fraction of this lifetime only. In addition, these massive stars are rare (less than 1~\% of the galactic stellar population) and therefore distant while the detection and characterization of exoplanets around these bright stars is very challenging with our current observation techniques. 
At the other end of the stellar population, M stars represent $\sim 75$~\% of the all stars, live tens to hundreds of Gy with a very stable luminosity (except for their earliest stage, which we will discuss below). Temperate terrestrial planets are frequent around these red-dwarfs and their characterization works far better with today's instruments. Therefore, along with Sun-like stars, they constitute a potential key population for habitable planets host-stars. To illustrate the ways in which a star may influence the habitability of its planets, we compare throughout the manuscript how the habitability of a planet varies -- everything else being kept fixed -- depending on whether it is orbiting a sun-like star or a late M-star. 
For the latter, we chose to use our closest neighbour, the star Proxima Centauri, as the archetype of a late M star (M5.5V, T\textsubscript{eff}~=~3090K) ; firstly because the star has been extremely well documented (see \cite{Ribas:2017} and references therein) ; secondly because with the detection of a terrestrial-mass planet in temperate orbit (\cite{Anglada:2016,Damasso:2020,Suarez:2020,Kervella:2020}), it presents one of the most promising systems for our quest to find a habitable planet outside the solar system (\cite{Turbet:2016,Lovis:2017,Boutle:2017,Leung:2020}). 

The manuscript is organized as follows.
Firstly, we review in Section~\ref{section_star_flux} how the spatial and spectral distribution of the incident bolometric (mostly visible and near-infrared) flux affects the habitability of planets. Secondly, we discuss in Section~\ref{section_UV_radiation} the influence of the far and mid-UV incident flux on the photochemistry, which can further affect the atmospheric composition and thus the habitability of planets. Thirdly, we present in Section~\ref{section_X-EUV_radiation} how the X and EUV incident radiation can affect atmospheric escape, which can not only affect atmospheric composition but also endanger the very existence of an atmosphere around planets, which has severe consequences for their habitability. Fourthly, we show in Section~\ref{section_temporal_evolution} how the temporal evolution of the bolometric emission of a star (as well as its most energetic part, e.g. the XUV emission) impacts the habitability of planets around it.

\section{Influence of the bolometric emission distribution on the climate}
\label{section_star_flux}

In this section, we briefly review how the climate and thus the habitability of a planet can be directly affected by the spectral type of its host star. The first naive effect is that the spectral type of the star changes its luminosity (for example, the luminosity of Proxima Cen is 0.15$\%$ that of the Sun) and thus the incident bolometric flux received by a planet at a fixed distance from the star. However, for a fixed amount of incident bolometric radiation received by the planet, the way in which stellar bolometric emission is distributed spectrally and spatially directly affects the way in which the surface, atmosphere and clouds absorb and reflect incident light, which changes the planet's surface temperature and thus the conditions of habitability.

\begin{figure*}
    \centering
\includegraphics[width=\linewidth]{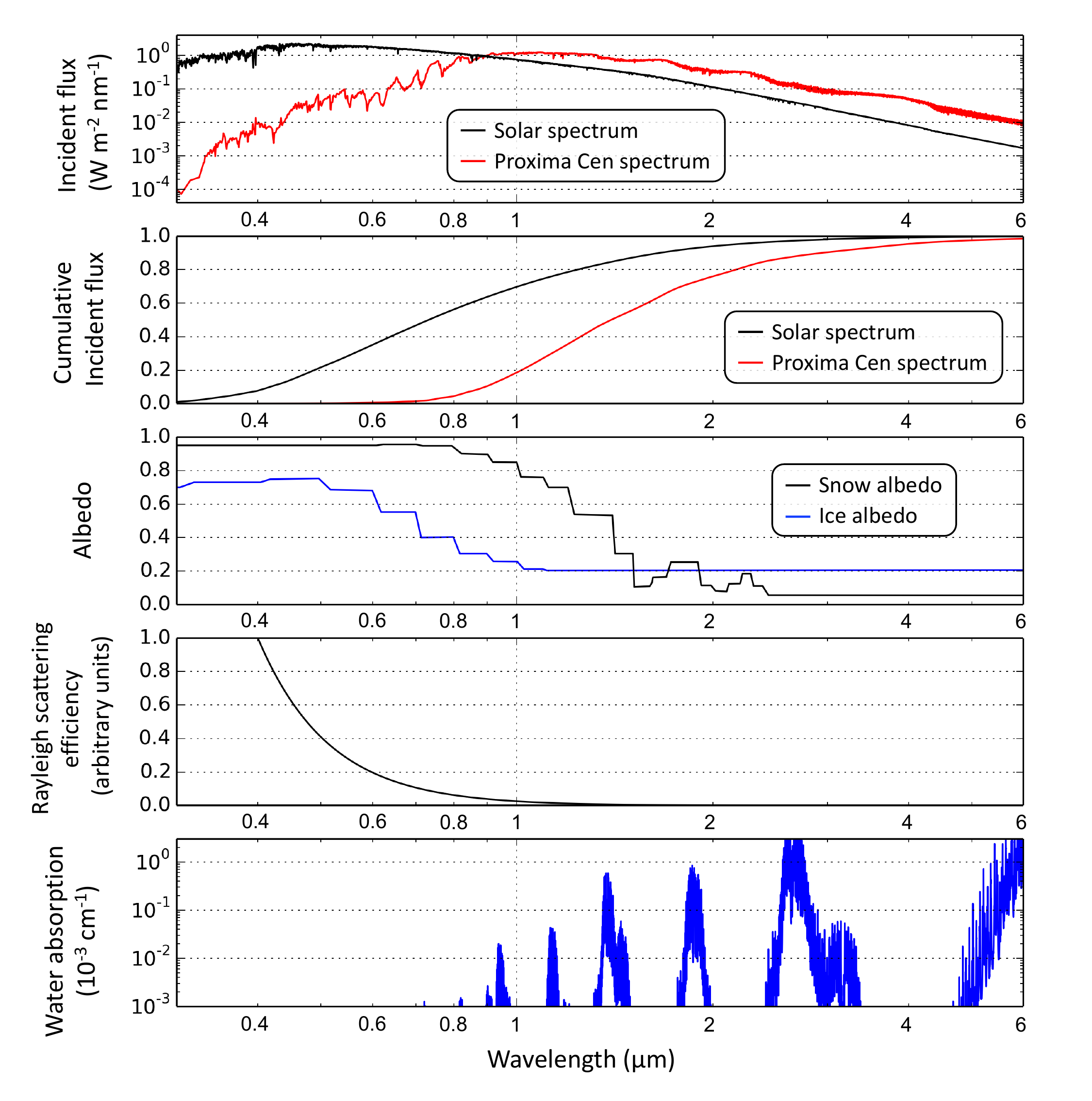}
\caption{\underline{First panel:} Incident flux spectra (at the top of the atmosphere) on a planet receiving a total of 1362~W~m$^{-2}$ (i.e. the solar constant on Earth ; note that the average flux received on the planet is four times lower, i.e. 340.5~W~m$^{-2}$). The black curve is based on solar spectra reconstructed from \cite{Meftah:2018} for $\lambda~<~$3$\mu$m and from \cite{Kurucz:1984} for $\lambda~\ge~$3$\mu$m. The red curve is based on Proxima Centauri spectra reconstructed from \cite{Ribas:2017}. \underline{Second panel:} Cumulative incident flux spectra. \underline{Third panel:} Albedo spectra for snow and water ice from \cite{Joshi:2012}. \underline{Fourth panel:} Rayleigh scattering cross-section spectrum (in arbitrary units). \underline{Fifth panel:} Water vapor absorption spectrum, calculated at 200~K, 1~bar, and 0.1$\%$ of water vapor.}
\label{spec_compare}
\end{figure*}

The main factor in the spectral distribution of the incident bolometric flux on a planet is the effective temperature of the host star. According to Wien's law, the cooler the star, the more the bolometric emission is shifted to long wavelengths. As a matter of fact, while the peak of the Sun's emission (T\textsubscript{eff}~=~5780K) is $\sim$~0.47~$\mu$m, that of Proxima Centauri (T\textsubscript{eff}~=~3090K) is $\sim$~1.1~$\mu$m (see Fig~\ref{spec_compare}a). Note that due to the spectral shape of the Planck's law, the median of the stellar emission is shifted to higher wavelengths (compared to the peak). Median of solar emission is $\sim$~0.73~$\mu$m while that of Proxima Centauri is $\sim$~1.4~$\mu$m (see Fig~\ref{spec_compare}b).

This has important consequences on a planet, because the surface, the atmosphere and the clouds absorb and/or scatter light differently at different wavelengths. Firstly, spectral variations of the surface albedo can have severe consequences on the stability of liquid water at a planetary surface. Ice and snow albedo are high at short wavelengths, and low at long wavelengths (see Fig~\ref{spec_compare}c). As a result, ice and snow are much more reflective around a sun-like star than around a low-mass star, indicating that planets orbiting solar-type stars are more prone to glaciation (e.g., to be trapped in a snowball state) than planets orbiting low-mass stars (\cite{Joshi:2012,Shields:2013,Vonparis:2013b}). More generally, spectral variations in surface albedo, which have now been documented for many types of surface (\cite{Hu:2012}), can have a significant effect on a planet's climate (\cite{Madden:2020}).
Secondly, Rayleigh scattering by the atmospheric gases can efficiently back-scatter (i.e. reflect back to space) incident stellar light, as illustrated by \cite{Keles:2018} for N$_2$-dominated atmospheres, or by \cite{Kopparapu:2013} for CO$_2$-dominated atmospheres. Given that the Rayleigh scattering efficiency is $\propto$~$\lambda ^{-4}$ (see Fig~\ref{spec_compare}d), the reflection of incoming stellar light by this process is much more efficient around a solar-type star than a low-mass star. 
Thirdly, direct absorption of incident stellar radiation by gases can vary greatly depending on the type of star. Most common gases lack strong absorption features in the optical wavelengths (\cite{Gordon:2017}). This is illustrated in Fig~\ref{spec_compare}e for H$_2$O absorption bands that absorb preferentially in the Proxima Cen emission wavelengths than those of the Sun. 
By combining the three ingredients mentioned above, we can conclude that -- for a given incident flux -- low-mass stars warm habitable planets more efficiently than sun-like stars do. This explains why the boundaries of the Habitable Zone are shifted towards lower incident stellar fluxes for low-mass stars (see e.g., Fig~8 in \cite{Kopparapu:2013}).

\begin{figure*}
    \centering
\includegraphics[width=\linewidth]{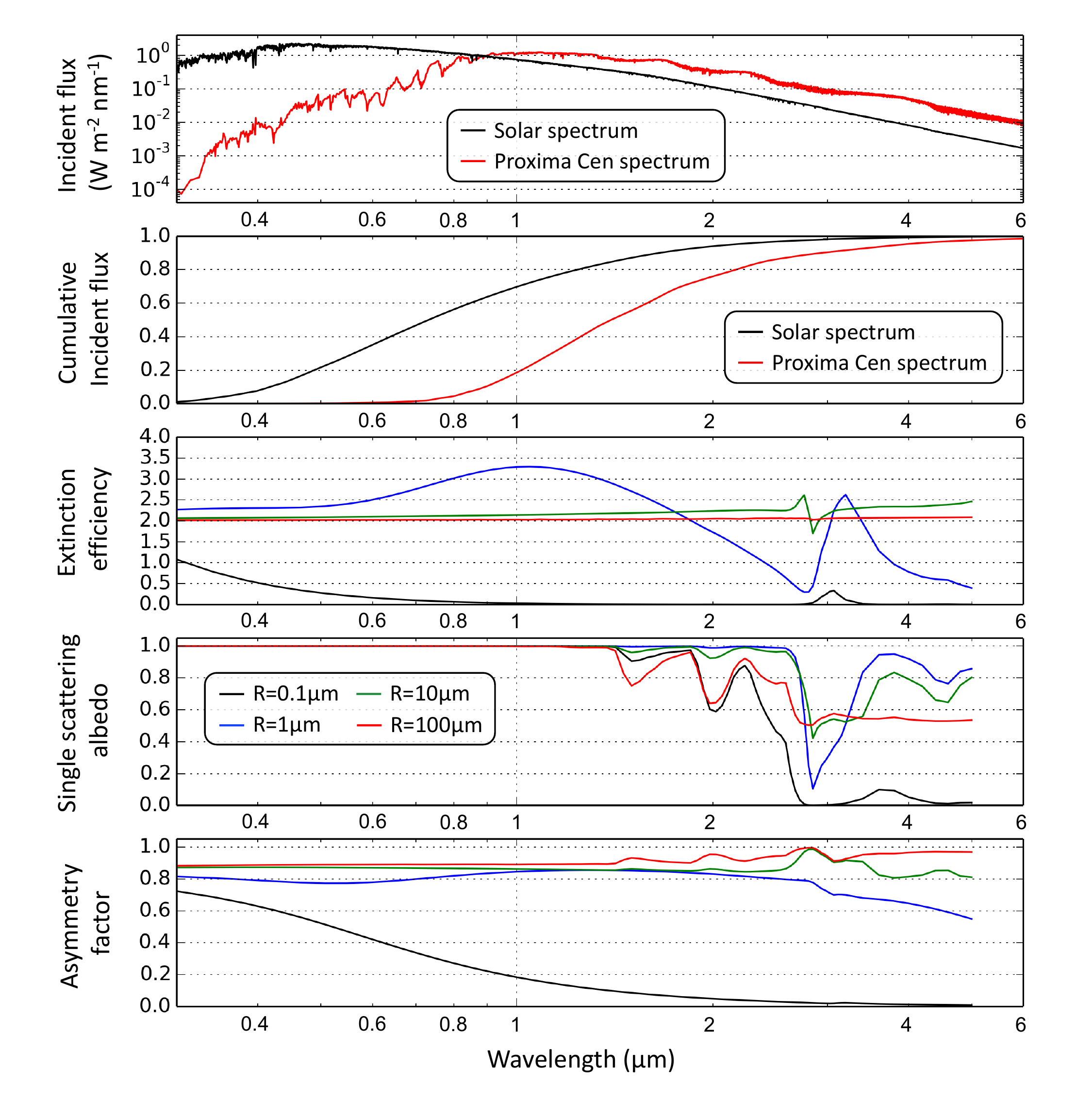}
\caption{\underline{First panel:} Incident flux spectra as in Fig.~\ref{spec_compare}, panel 1. \underline{Second panel:} Cumulative incident flux spectra as in Fig.~\ref{spec_compare}, panel 2. \underline{Third panel:} Water ice cloud extinction efficiencies for 4 spherical particle sizes (radius of 0.1, 1, 10 and 100$\mu$m). \underline{Fourth panel:} Water ice clouds single scattering albedos (i.e., the ratio of scattering efficiency to total extinction efficiency ; single-scattering albedo is equal to 1 if all is due to scattering, and equal to 0 of if all particle extinction is due to absorption). \underline{Fifth panel:} Water ice clouds asymmetry factors (i.e., the mean cosine of the scattering angle ; the asymmetry factor is 0 for isotropic radiation and ranges from -1 [negative values mean backward scattering is predominant] to 1 [positive values mean forward scattering is predominant].)}
\label{spec_clouds}
\end{figure*}

Clouds may also have different properties whether they are exposed to a solar-like or a low-mass star incident radiation. This is illustrated in Figs.~\ref{spec_clouds}c, d and e which show the spectra of the key microphysical properties (extinction efficiency, single scattering albedo and asymmetry factor) of water ice clouds for several particle radii (\cite{Fu:2006}).
Whether the presence of a particular type of ice particle or liquid water droplet serves to increase or decrease planetary albedo depends on the interplay among extinction efficiency, single scattering albedo and asymmetry factor (\cite{Fu:2006}).
Clouds made of small particles (e.g., 0.1~$\mu$m) are very reflective (asymmetry factor close to 0, single scattering albedo close to 1) for a wide range of stellar types. While clouds made of larger particles (e.g., 1~$\mu$m and larger) are still highly reflective, they can absorb an appreciable fraction of the incident flux near 1.5, 2 and 2.9$\mu$m, which favors warming for planets around low-mass stars. However, overall, it has been shown that the differences between the different types of stars should be relatively small (\cite{Kitzmann:2010}).

\begin{figure*}
    \centering
\includegraphics[width=\linewidth]{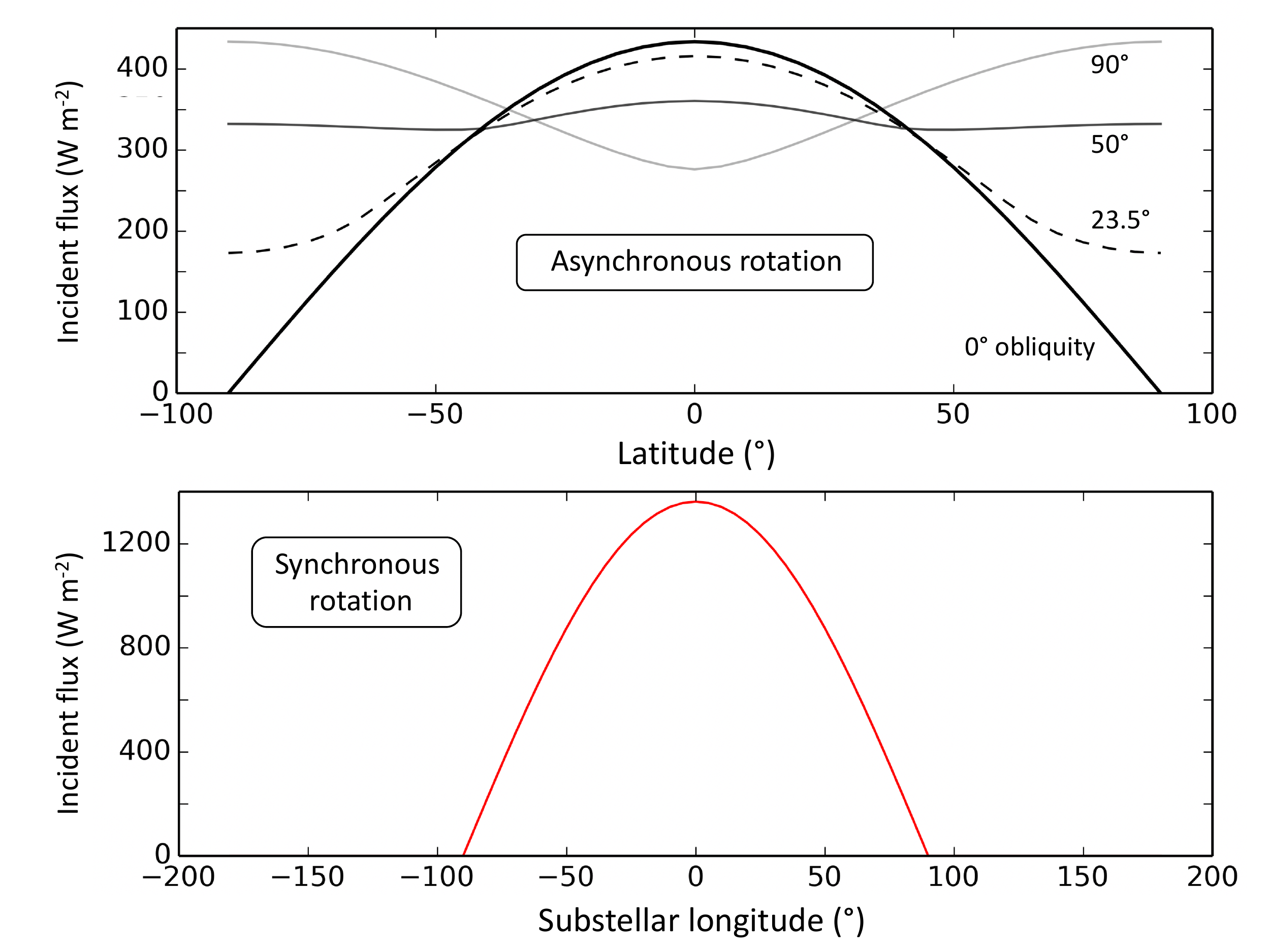}
\caption{\underline{First panel:} Zonal annual mean average of the incident flux on an asynchronously-rotating planet receiving a total of 1362~W~m$^{-2}$ (i.e. the solar constant on Earth), for four distinct obliquities (0, 23.5, 50, 90$^{\circ}$). \underline{Second panel:} Incident flux on a synchronously-rotating planet (also receiving a total of 1362~W~m$^{-2}$) versus substellar longitude. The zonal fluxes were calculated using equation~1 of \cite{Nadeau:2017}.}
\label{distrib_insolation}
\end{figure*}

The main effect by which clouds may affect the climates of habitable planets differently depending on the type of host star is actually related to their spatial distribution. It has indeed been shown that planets orbiting in temperate orbits around low mass stars are prone to be locked in a state of synchronous rotation (\cite{Dole:1964,Kasting:1993,Barnes:2017}), for which one planet hemisphere is permanently facing its host star. This has severe consequences for the distribution of the insolation at the TOA (top of atmosphere) or surface of a planet, as illustrated in Fig.~\ref{distrib_insolation}. While the solar flux is on average relatively evenly distributed on a fast-rotating planet (in particular at large obliquities ; see Fig.~\ref{distrib_insolation}a), the stellar flux received on a synchronously-rotating planet is highly concentrated on one side of the planet (see Fig.~\ref{distrib_insolation}b).
It has been shown that on a planet covered with liquid water, this concentration of stellar flux leads to strong moist convection concentrated in the substellar region, which leads to the formation of a thick layer of reflecting clouds (\cite{Yang:2013,Kopparapu:2016,Yang:2019,Fauchez:2020gmd}). These clouds, because of their stabilizing effect (higher temperature means stronger convection, which means more clouds are formed, which means incoming stellar radiation is reflected more efficiently, which cools the planet), have a first-order effect on the climate and habitability of synchronously-rotating planets (\cite{Yang:2013,Kopparapu:2016}), and thus on a significant fraction of planets orbiting low mass stars (\cite{Leconte:2015}).

\section{Influence of the UV emission on the photochemistry}
\label{section_UV_radiation}

The stellar Hydrogen Lyman-$\alpha$ (at 121.6~nm), far-ultraviolet ($\sim$~122-200~nm) and mid-ultraviolet ($\sim$~200-300~nm) spectral emissions are the main drivers of planetary photochemistry. UV ($\sim$~100-300~nm) photons can in fact photolize (i.e. break) atmospheric molecules, forming radicals capable of generating multiple, complex chemical reactions. These photochemical reactions can deeply affect the atmospheric composition of a planet in a way that depends on initial composition and the total amount and spectral distribution of the UV stellar emission (\cite{Yung:1999}). Eventually, the photochemically-driven destruction or buildup of greenhouse gases in the atmosphere will affect planetary surface temperature and thus the surface habitability of the planet.

\begin{figure*}
    \centering
\includegraphics[width=\linewidth]{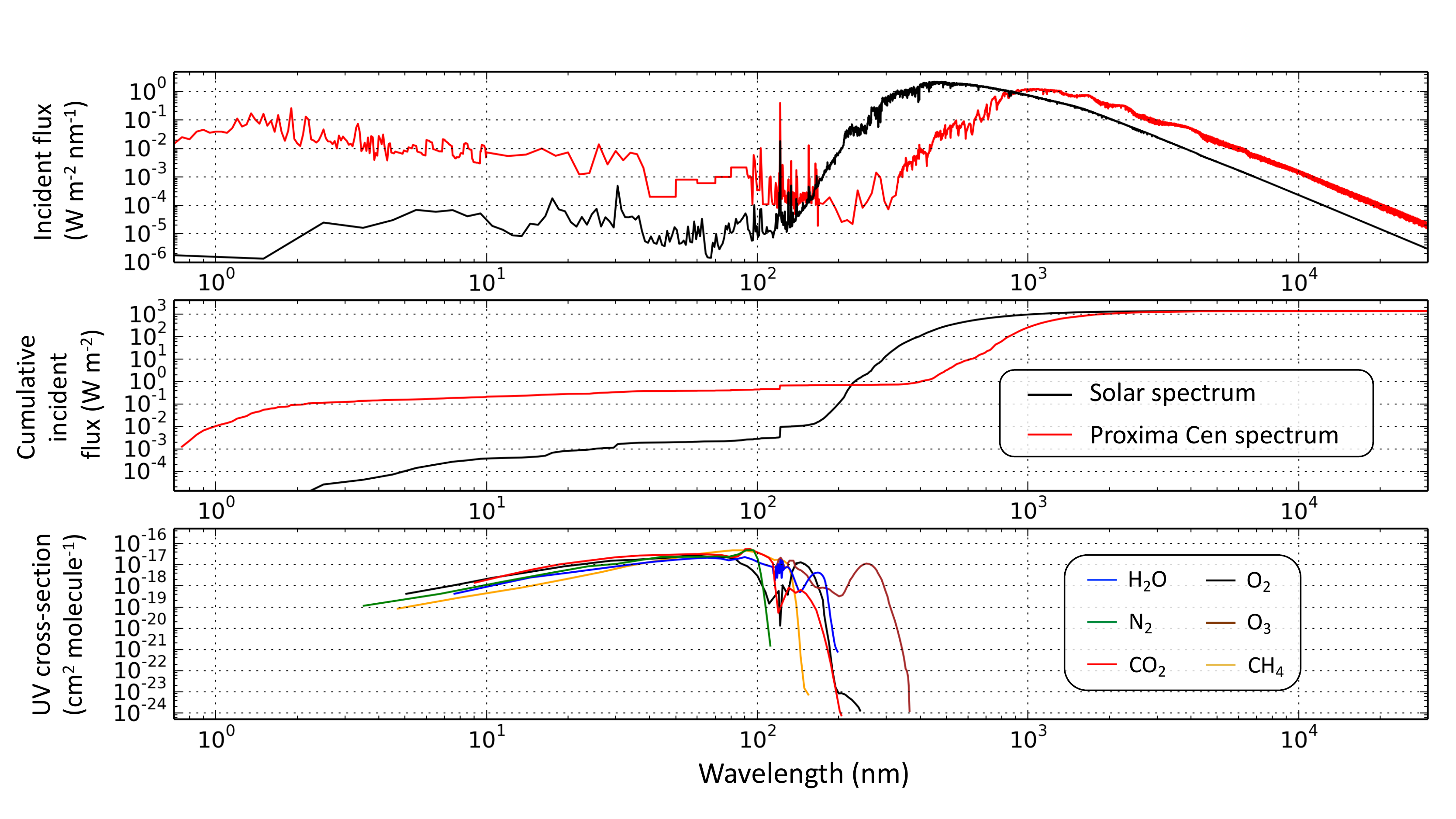}
\caption{\underline{First panel:} Incident flux spectra (at the top of the atmosphere) on a planet receiving a total of 1362~W~m$^{-2}$ (i.e. the solar constant on Earth). The black curve is based on solar spectra reconstructed from \cite{Meftah:2018} for $\lambda~<~$3$\mu$m and from \cite{Kurucz:1984} for $\lambda~\ge~$3$\mu$m. The red curve is based on Proxima Centauri spectra reconstructed from \cite{Ribas:2017}. \underline{Second panel:} Cumulative incident flux spectra. \underline{Third panel:} UV cross-sections for common atmospheric species (H$_2$O, N$_2$, CO$_2$, O$_2$, O$_3$, CH$_4$) in terrestrial planetary atmospheres. The cross-sections were taken from \cite{Keller-Rudek:2013}.}
\label{spec_uv_xsec}
\end{figure*}

Each molecule has an UV absorption cross-section (see Fig.~\ref{spec_uv_xsec}c) that, combined with the quantum yield (i.e. the probability that an absorbed photon actually breaks the molecule) and the UV incident spectral flux (Fig.~\ref{spec_uv_xsec}a), gives the photolysis rate of the molecule. Depending on the spectral type of the host star, the photolysis rate of atmospheric molecules may -- everything else being kept fixed -- significantly vary. For illustration, the flux received by a planet (given a fixed bolometric incident flux) around a low-mass star is 2-3 orders of magnitude lower than that of a planet around a solar-type star in the main spectral region for ozone photodissociation (Hartley band ; near $\sim$~250~nm), a clue that oxygen and ozone chemistry must operate in a very different way around these two stars. More generally, the flux dichotomy near $\sim$~150~nm (relative to its total bolometric luminosity, a M-star emit more photons below 150nm, and much less photons above 150nm) should have strong impact on the photochemistry of planetary atmospheres (\cite{Arney:2017,Meadows:2018proxima}). 

Although in most atmospheres the impact of photochemistry is primarily to modify the relative abundances of minor species (\cite{Rugheimer:2015,Arney:2017,Meadows:2018proxima}), there are some situations where photochemistry can entirely change the composition of an atmosphere (\cite{Titan:1997science,Gao:2015,Luger:2015,Hu:2020}). 

Finally, photochemistry can also lead to the formation of long carbonated chains suspended in the atmosphere, namely hazes. These radiatively active photochemical hazes can deeply affect the climate, either directly by absorbing and reflecting incoming stellar radiation in particular through Mie scattering, often resulting in surface cooling, or either indirectly through feedbacks on photochemistry e.g. by shielding molecules from incident UV radiation (\cite{Arney:2017}).

\section{Influence of the X-EUV emission on atmospheric loss}
\label{section_X-EUV_radiation}

X ($<$~10~nm) and Extreme-UV ($\sim$~10-100~nm) stellar emissions, also referred together as XUV, should play a fundamental role in atmospheric escape (\cite{Lammer:2003,Tian:2015c,Catling:2017}). XUV radiation heats the planetary thermosphere up to the exobase, which corresponds to the boundary where the atmosphere transitions from a collisional to a collisionless region. Hydrogen atoms are light enough to be lost to space when reaching the exobase even for a low XUV heating, like that of the Earth. The erosion of the water reservoir through this process is thus not controlled by the XUV flux but by the upward transport and photolysis of H$_2$O. Thanks to the cold trap of the tropopause that results in a H$_2$O-poor stratosphere, the erosion of Earth's water reservoir is negligible. When the XUV flux is sufficiently high, models predict that the heating powers an hydrodynamic expansion of the atmosphere, which is lost through a planetary wind (\cite{Johnstone:2019b}). Even if hydrogen is the only species likely to be lost hydrodynamically with realistic XUV fluxes, other heavier species (e.g., oxygen) can also be accelerated through collisions and then be lost to space (\cite{Johnstone:2020}). In this hydrodynamical regime, the loss rate is controlled by the intensity of the XUV flux (as well as the availability of hydrogen atoms).

XUV incident flux on a planet can strongly vary depending on the type of host star, e.g. between 2-5 orders of magnitude depending on the exact wavelength between solar-type and low-mass stars (see Fig.~\ref{spec_uv_xsec}a ; for a fixed bolometric incident flux), and depending also on the level of activity of the low-mass star. The total XUV flux of Proxima Cen is ~$\sim$~200 times higher than that of the Sun (see Fig.~\ref{spec_uv_xsec}b), relatively to their total bolometric emission. This means that the atmospheric escape rate can be significantly higher for planets around late M-stars than for Sun-like stars.

Most of stellar-driven atmospheric escape processes including XUV-driven escape (but also driven by the stellar wind drag or the photochemical escape) are also expected to be stronger around low-mass stars than solar-type stars (again, for a given total bolometric flux).

This is important information, given atmospheric escape can change the chemical composition or remove most of all of a planetary atmosphere. In the future, this threat against habitability around active late stars will have to be assessed with both observations/statistics of planetary atmospheres in various XUV/particle irradiation contexts and with robust modeling of the escape processes. The latter will require significant progress in order to treat in 3-D the transition between the fluid expanding atmosphere and the diluted particle regime where kinetic and electromagnetic interactions occurs far from LTE.

\section{Influence of the temporal evolution of the stellar emissions}
\label{section_temporal_evolution}

From birth to death, stars evolve in luminosity. They start their life with a super-luminous pre-main sequence phase during which they cool by radiating their initial accretion energy, then evolve on the main sequence during which their luminosity gradually increases as they fuse hydrogen into helium (\cite{Chabrier:1997,Baraffe:1998,Baraffe:2015}) and until they finally die.
The duration of each of these two phases (pre-main sequence and main sequence), as well as the evolution of luminosity across them, has major consequences on the climate and habitability of the planets.

Firstly, the duration of the main sequence phase, which is all the longer the smaller the star, plays a key role in the stability of habitability conditions over time. While the duration of the main sequence of the Sun is a few billion years (but can be much shorter for higher mass stars), that of very low mass stars is expected to be much longer than the age of the Universe.

\begin{figure*}
    \centering
\includegraphics[width=\linewidth]{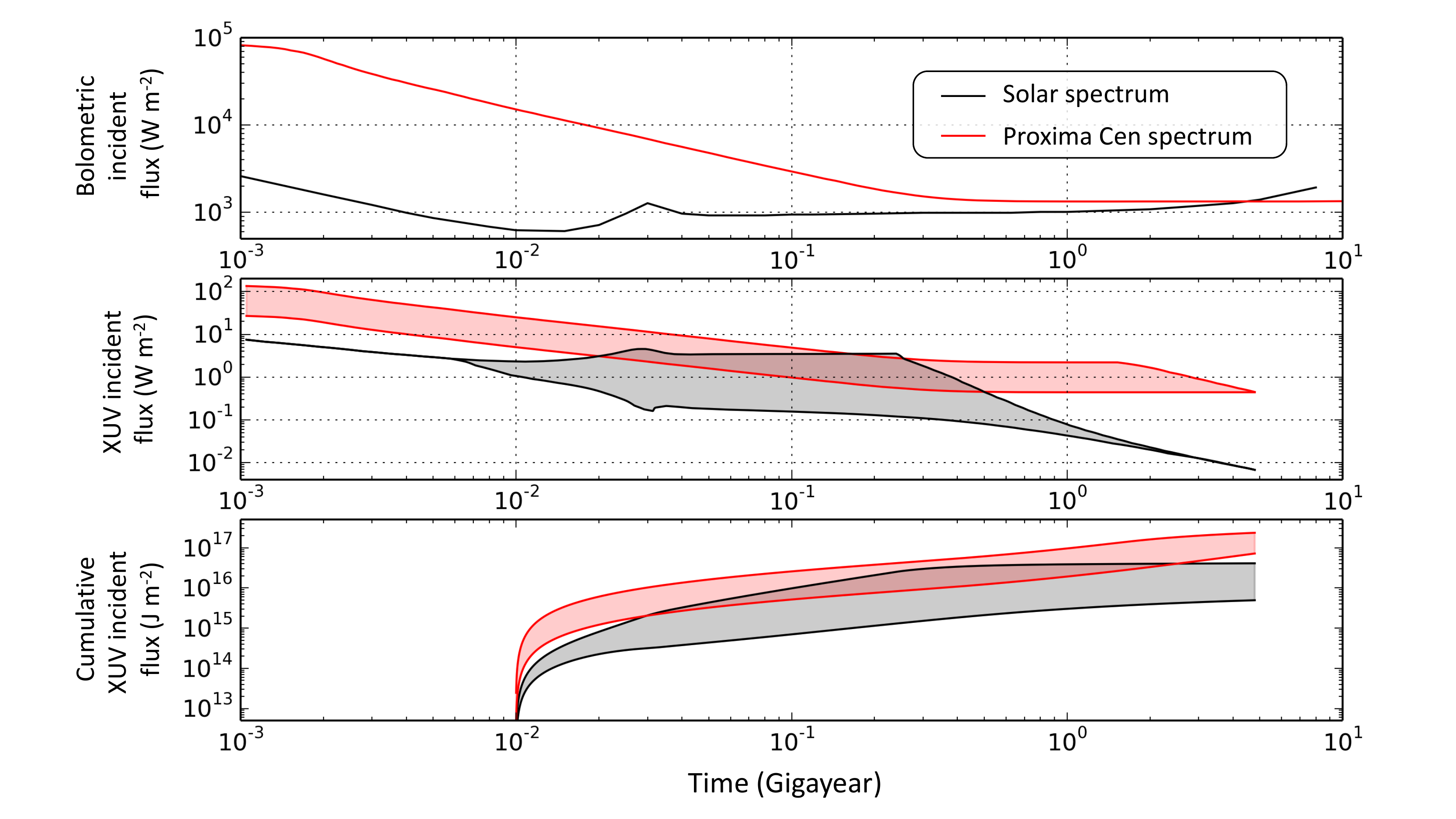}
\caption{\underline{First panel:} Temporal evolution of the incident flux (at the top of the atmosphere) on a planet receiving a total of 1362~W~m$^{-2}$ (i.e. the solar constant on Earth). The black curve is based on the solar evolution model of \cite{Baraffe:2015}. The red curve is based on the Proxima Centauri evolution model of \cite{Ribas:2017}. \underline{Second panel:} Temporal evolution of the XUV (10-120~nm) incident flux (at the top of the atmosphere) on a planet receiving a total of 1362~W~m$^{-2}$ (i.e. the solar constant on Earth). The black curves are based on the XUV emission solar models of \cite{Tu:2015} (based on \cite{Gallet:2013}) for slow (lower curve) and fast (upper curve) rotator scenarios. The red curves are based on the two evolutionary scenarios of \cite{Ribas:2017}: (1) Proxima Cen has spent its entire lifetime in the XUV emission saturation regime (lower curve); (2) Proxima Cen was in the saturation regime only for the first $\sim$1.6~Gigayear of evolution (now aged of $\sim$~4.8~Gigayear). \underline{Third panel:} Cumulative XUV incident flux starting 10~million years after the star formation, which is around the time the planets are expected to have formed.}
\label{evol_luminosity}
\end{figure*}

Secondly, while the pre-main sequence phase duration of solar-type stars is very short ($\sim$~10~million years), that of very low mass stars can last for several hundreds of million years. During this phase, planets are exposed to a much larger incident flux, up to 10~x higher for planets around Proxima Cen (Fig.~$\ref{evol_luminosity}$a). Planets that receive today the right incident flux to maintain surface liquid water were likely so hot (in the pre-main sequence phase of their host star) that they experienced a runaway greenhouse (\cite{Ramirez:2014c}), with a thick steam atmosphere (assuming they formed with significant amounts of water) but no surface liquid water.

This is particularly critical for the habitability of planets around such stars, because water in the form of vapor is exposed to strong loss to space (\cite{Luger:2015,Bolmont:2017,Bourrier:2017b,Wordsworth:2018,Johnstone:2020}). Pre-main sequence phase can in fact potentially lead to complete water loss on a planet, depending on the initial reservoir. This is even more critical knowing that XUV heating, i.e. the main process by which the planet is likely to lose its atmosphere and its water, is also much more efficient during the pre-main sequence phase. This stems from the fact that the XUV radiation decreases even more rapidly than bolometric radiation during the pre-main-sequence phase (see Fig.~$\ref{evol_luminosity}$b). Note that while the XUV flux emitted by a solar-like star may also have been very high in the past (at least for the fast rotator population ; see \cite{Tu:2015}), the cumulative XUV radiation (see Fig.~$\ref{evol_luminosity}$c) is in general significantly larger around low-mass stars than sun-like stars.
Combined with the long pre-main-sequence phase during which the planets are more irradiated than the runaway greenhouse threshold, and therefore during which water is exposed to atmospheric escape, this indicates that water loss is potentially much stronger on planets around late M-stars than around sun-like stars, which can have serious consequences for their habitability.

\section{Conclusions}
\label{section_conclusions}

We briefly reviewed the main effects by which a star can have an impact on the habitability of planets in orbit around it. Firstly, spectral and spatial variations of the bolometric (i.e. visible and near-infrared) stellar emission can affect the way a planet's surface, atmosphere and clouds will absorb or reflect incident light. These changes can affect the climate of a planet (especially the surface temperature) which can thus impact its habitability. 
Secondly, variations in UV (far and mid-UV) incident fluxes on a planet can drive different photochemistry. This can potentially cause significant changes in the atmospheric composition, and thus the radiation balance of a planet, which can again strongly impact its habitability.
Thirdly, cumulative variations in XUV (X and extreme-UV) incident fluxes on a planet can lead to various degrees of atmospheric erosion, potentially leading to dramatic effects (e.g. through the partial or complete loss of the atmosphere and the water) on the habitability.

Taking into account at the same time all the effects by which a star interacts with its planets in the same model, to simulate in a coherent way the evolution of the atmosphere and the habitability of the planets around any type of star, is one of the great challenges of the field.  This ambitious scientific objective requires, among other things, a precise knowledge of the spectral properties (from infrared to XUV radiation) of the star at the time it is observed, but also of how these properties have evolved over time.

\section*{Acknowledgements}
This project has received funding from the European Union’s Horizon 2020 research and innovation program under the Marie Sklodowska-Curie Grant Agreement No. 832738/ESCAPE. M.T. thanks the Gruber Foundation for its support to this research. This work has been carried out within the framework of the National Centre of Competence in Research PlanetS supported by the Swiss National Science Foundation. The authors acknowledge the financial support of the SNSF.


\bibliographystyle{apalike} 
\bibliography{biblio} 

\end{document}